\documentstyle[aps]{revtex}

\begin{document}
\title{Complex version KdV equation and the periods solution}
\author{Yang Lei, Yang Kongqing, Luo Honggang}
\address{({\it Department of physics, Lanzhou University, Gansu 730000, China)}}
\date{29/6 1999}
\maketitle

\begin{abstract}
In this paper, the complex version KdV equation is discussed. The
corresponding coupled equations is a integrable system in the sense of the
bi-Hamiltonian structure, so the complex version KdV equation is integrable.
A new spectral form is given, the periodic solution of the complex version
KdV equation is obtained. It is showed that the periodic solution is the
classical solution.
\end{abstract}

\section{Introduction}

Since 80's, the coupled KdV equations as an important mathematical model has
been studied widely. In 1981, Fuchssteiner[1] made a detailed study of four
coupled KdV equations and gave the bi-Hamiltonian structure of them. One
coupled KdV equations is the complexly-coupled KdV equations

\begin{equation}
\begin{array}{l}
u_t=u_{xxx}+6uu_x+6\varphi \varphi _x, \\ 
\varphi _t=\varphi _{xxx}+6u\varphi _x+6u_x\varphi .
\end{array}
\label{1}
\end{equation}
Later, Oevel[2] pointed out that ''......inserting a complex ansatz $%
u+i\varphi $ into the KdV......it is a complex version of the KdV......''
and the complex version of the KdV poses two conservation laws in every
order. On the following day, Hirota and Satsuma introduced a coupled KdV
equations[3], the integrability of the equations was discussed by the
bi-Hamiltonian structure[1,2] and Lax pair[4]. Recently, more coupled KdV
equations and their bi-Hamiltonian structure has been presented [5,6].

In this paper, the complex version KdV equation is considered,

\begin{equation}
u_t+\alpha uu_x+\beta u_{xxx}=0,  \label{2}
\end{equation}
where $x$ is a one-dimension space, $t$ is a one-dimension time, $\alpha $
and $\beta $ are coefficients, which play important role in our numerical
results, $u(x,t)=p(x,t)+iq(x,t)$ is an unknown variable, $p(x,t)$ is the
real part of $u(x,t)$, $q(x,t)$ is the imaginary part of $u(x,t)$. Extending
the lax pair of the KdV equation to complex field, $L=-D^2+p+iq$, $%
M=6(p+iq)D-4D^3+3p_x+3iq_x$, the complex version KdV equation can be derived
from the lax equation $L_t=\left[ M,L\right] $. And the complex lax pair
binds with the complex potential scattering and inverse scatting problem of
the schr\H{o}dinger equation, which had been studied in a optical
problem[7]. Here, let $\alpha =\beta =1$, substituting the $%
u(x,t)=p(x,t)+iq(x,t)$ into the equation(2), separating the real part and
the imaginary part from the equation(2), a coupled KdV equations is obtained

\begin{equation}
\begin{array}{l}
p_t+p_{xxx}+pp_x-qq_x=0, \\ 
q_t+q_{xxx}+pq_x+p_xq=0.
\end{array}
\label{3}
\end{equation}
The equations(3) are equal to the equation(2) completely and has a little of
differences with the Fuchssteiner's equation(1). Under the transformation: $%
p\rightarrow 6u$, $q\rightarrow 6i\phi $, $x\rightarrow -x$, the equation(3)
are exactly equivalent to equation(1).

\section{Method describing and the periodic solution}

In general, the nonlinear evolution equation is written as $%
u_t(x,t)=F(u(x,t),u_x(x,t),\ldots )$, here $F(u)$ is a non-linear function
of $u(x,t),u_x(x,t),\ldots $ , which is a nonlinear operator involving only
the spatial derivatives. Take the Fourier expansion,

\begin{equation}
u(x,t)=\sum\limits_{n=-\infty }^\infty u_n(t)e^{ik_nx}  \label{4}
\end{equation}
where $k_n=2n\pi /L$ and $L$ is a period interval in space. Let expression
(4) come in with $u_t(x,t)=F(u(x,t))$, receive the infinite reference
ordinary differential equations(ODE), which is highly nonlinearity and
complexity in nature, the exact solutions of the ODEs can't be obtained.

In this paper, a new spectral form is given, expanding $u(x,t)$ on the
sub-set of Fourier basis \{$%
\mathop{\rm e}%
^{ik_nx},n=0,1,2,...\}$, namely,

\begin{equation}
u(x,t)=\sum\limits_{n=0}^\infty u_n(t)e^{ik_nx}  \label{5}
\end{equation}
where $k_n=2n\pi /L$. Substitute the formula (5) into $u_t(x,t)=F(u(x,t))$,
thus 
\begin{equation}
\sum\limits_{n=0}^\infty \frac{du_n}{dt}e^{ik_nx}=F(u).  \label{6}
\end{equation}
Multiplication of equations (6) by $e^{-ik_mx}$, and integration over a
period $2\pi /L$ gives 
\begin{equation}
\sum\limits_{n=0}^\infty a_{mn}\frac{du_n}{dt}=p_m(u),  \label{7}
\end{equation}
where $a_{mn}=\delta _{mn}$, $p_m(u)=\frac L{2\pi }\int_0^{2\pi
/L}F(u)e^{-ik_mx}\,dx$. Apparently, the equation(5) with (6) and (7) reduces
to a ordinary differential equations, 
\begin{equation}
\frac{du_m}{dt}=p_m(u)~~~~~~~~for~~~m=0,1,2,\cdots ,  \label{8}
\end{equation}
where $p_m$ is a function of the $u_m$ , for the KdV equation $F(u)=-[\alpha
uu_x+\beta u_{xxx}],$ the (8) can be worked explicitly out, 
\begin{equation}
\frac{du_m(t)}{dt}=i\alpha \sum\limits_{\stackrel{k,l=0}{k+l=m}%
}^mu_k(t)lu_l(t)\ -i\beta m^3u_m(t)\ \ \ \ for\ \ \ m=0,1,2,\cdots .
\label{9}
\end{equation}
Therefore, underlying the basis (4) the KdV equation becomes to a hierarchy
of infinite first-order ordinary differential equations. It is easy to solve
these ordinary differential equations by the standard integral method,

\begin{equation}
u_m(t)=e^{-i(-m\alpha u_0+\beta m^3)t}[\int i\alpha \sum\limits_{\stackrel{%
k,l=1}{k+l=m}}^{m-1}u_k(t^{\prime })lu_l(t^{\prime })e^{\ i(-m\alpha
u_0+\beta m^3)t^{\prime }}\,dt^{\prime }+C_m],  \label{10}
\end{equation}
where $C_m$ is integration constant, and $u_0$ also any constant from$\frac{%
du_0(t)}{dt}=0.$ When taking $C_1=1$, $C_m(m\geq 2)$ for zero,

\begin{equation}
\begin{array}{l}
u(x,t)=\sum\limits_{n=1}^\infty c_ne^{int}e^{-inx}, \\ 
c_n=\frac 6{n(n^2-1)}\sum\limits_{\stackrel{k,l=1}{k+l=n}%
}^{n-1}c_klc_l~~~~~~~(n\geq 2),
\end{array}
\label{11}
\end{equation}
here $c_1=1$, $c_n$ is the coefficient. Although the base of the sub-space \{%
$e^{ik_nx},n=0,1,2,...\}$ is incomplete, it is closed under the operation of
the KdV equation. Under the closed condition and Sobolev theorem of
embedding, one can show that the solution is the classical solution of the
corresponding equation(Appendix A and B give the proof).

\section{Conclusion}

In conclusion, based on the incomplete basis $\{e^{-inx},n=0,1,2,\cdots \}$,
the complex version KdV equation(2) is discussed. Under the operation of the
KdV equation, the basis is closed and forms an invariant set. The classical
periodic solution was obtained in this invariant set. In Appendixes, it is
showed that this solution is the classical solution of the KdV equation.

\section{Appendix A: The proof of the convergence of the periodic solution}

The convergence of the periodic solution

\[
\begin{array}{l}
u(x,t)=\sum\limits_{n=1}^\infty c_ne^{int}e^{-inx}, \\ 
c_n=\frac 6{n(n^2-1)}\sum\limits_{\stackrel{k,l=1}{k+l=n}%
}^{n-1}c_klc_l~~~~~~~(n\geq 2),
\end{array}
\]
here $c_1=1$, is discussed as follows: Considering the convergence of module
of coefficients. It is obvious that 
\[
0\leq |c_n|\leq \frac 6{n(n^2-1)}\sum\limits_{\stackrel{k,l=1}{k+l=n}%
}^{n-1}|c_k|l|c_l|\leq 1~for~~n=1,2,\cdots .
\]
Then for arbitrary given $0<\epsilon <1/2$, one can show inductively that $%
|c_n|\leq \frac A{n^{3/2+\epsilon }}$, where $A\geq 1$ is a constant
depending on $\epsilon $. Note that 
\[
S_\epsilon =\sum\limits_{n=1}^\infty \frac 1{n^{3/2+\epsilon }}<\infty 
\]
and inequality

\[
\sum\limits_{k=1}^n\frac 1{k^\alpha }\leq \sum\limits_{k=1}^n\int_k^{k+1}%
\frac 2{x^\alpha }\,dx=2\int_1^{n+1}\frac 1{x^\alpha }\,dx\leq \frac{%
2(n+1)^{1-\alpha }}{1-\alpha }. 
\]
Without loss of generality, take $A\geq 64$. it is obvious that

\[
\begin{array}{c}
B_n=\frac{6n^{2(3/2+\epsilon )}}{n(n^2-1)}\{\sum\limits_{l=1}^{[\frac{n-1}2]}%
\frac 1{l^{1/2+\epsilon }(n-l)^{3/2+\epsilon }}+\sum\limits_{l=[\frac{n-1}2%
+1]}^{n-1}\frac 1{l^{1/2+\epsilon }(n-l)^{3/2+\epsilon }}\} \\ 
\leq \frac{6n^{2(3/2+\epsilon )}}{n(n^2-1)}\{\frac{2^{3/2+\epsilon }}{%
n^{3/2+\epsilon }}\sum\limits_{l=1}^{n-1}\frac 1{l^{1/2+\epsilon }}+\frac{%
2^{1/2+\epsilon }}{n^{1/2+\epsilon }}\sum\limits_{l=1}^{n-1}\frac 1{%
(n-l)^{3/2+\epsilon }}\} \\ 
\leq \frac{6n^{2(3/2+\epsilon )}}{n(n^2-1)}\{\frac{2^{5/2+\epsilon
}(n+1)^{1/2-\epsilon }}{(1/2-\epsilon )n^{3/2+\epsilon }}+\frac{%
2^{1/2+\epsilon }S_n}{n^{1/2+\epsilon }}\}.
\end{array}
\]
where $B_n=\frac{6n^{2(3/2+\epsilon )}}{n(n^2-1)}\sum\limits_{l=1}^{n-1}%
\frac 1{l^{1/2+\epsilon }(n-l)^{3/2+\epsilon }}$. It is obvious that $%
B_n\rightarrow 0$ at $n\rightarrow +\infty $, so there exists $n_0$ to make $%
B_n\leq 1$ at $n\geq n_0$. Thus take $A=max\{64,n_0^{3/2+\epsilon }\}$, for
given $A$. It is natural to draw a conclusion that the periodic solution
(11) is convergent.

\section{Appendix B: The proof of the classical solution}

The first proof, the solution(11) is weak solution.

For $0\leq t_1<t_2\leq T,~~Q_{t_1,t_2}={\bf R}\times [t_1,t_2]$. Definition: 
$u\in C(Q_T)$ is periodic weak solution with periodicity $P$ of the KdV
equation, if $u$ satisfy 
\[
\int_{t_1}^{t_2}\int_0^P\{u(-\varphi _t+\alpha \varphi _{xx}-\beta \varphi
_{xxx})-\frac 12u^2\varphi _x\}dxdt=0,
\]
where $\varphi (x,t)\in C^\infty (Q_{t_1,t_2}),\varphi (\cdot ,t_1)=\varphi
(\cdot .t_2)=0$, and $\varphi $ is a periodic function with periodicity $P$
for variable $x$. The following we take $P=2\pi $. Assuming $%
R_N(x,t)=u(x,t)-\sum\limits_{n=0}^Nu_n(t)e^{-inx}$, from Appendix A, it
reads $R_N(x,t)\rightarrow 0,~$as$~N\rightarrow \infty $. Definition $I_N$
is 
\[
I_N=\int_{t_1}^{t_2}\int_0^{2\pi }\{\varphi (D_tu_N+\alpha D_{xx}u_N+\beta
D_{xxx}u_N+\frac 12D_x(u_N^2))\}dxdt,
\]
where subscript $x,t$ denote differential, $u_N(x,t)=\sum%
\limits_{n=0}^Nu_n(t)e^{-inx}$. Apparently, $I_N$ tends zero as $%
N\rightarrow \infty $. Thus it's show that the solution of the KdV equation
is a weak solution by definition.

The second proof, the weak solution is classical solution.

Note the solution(11), following two conclusions have been showed:

(i),$~\forall A>0,0<\epsilon <\frac 12$, $|c_n|\leq \frac A{n^{\frac 32%
+\epsilon }}(n=1,2,\cdots )$.

(ii), For arbitrary $\eta \in C_0^\infty (-\infty ,\infty ),\phi (x,t)\in
C^\infty ({\bf R}^2)$, where $\phi (x,t)$ is a periodic function with
respect to the variable $x$, $u(x,t)$ satisfies 
\[
\int_{-\infty }^\infty dt\int_0^{2\pi }\{u(x,t)[-(\eta ^2\phi )_t+\eta
^2(\alpha \phi _{xx}-\beta \phi _{xxx})]-\frac 12\eta ^2u^2\phi _x\}dx=0 
\]
Based on the conclusions (i) and (ii), a standard step can be applied, it is
easy to show that the formal solution (11) is infinitely differentiable and
must be the classical solutions of the KdV equation at $t>0$. (The similar
conclusion can be draw at $t<0$.)

\section{ACKNOWLEDGMENTS}

The author is grateful to Professors Ya-Zhe Chen and Ming-Liang Wang for
valuable discussions.

The corresponding author email address: yangkq@lzu.edu.cn (Yang Kongqing).

\end{document}